\documentclass[twocolumn]{webofc}
\usepackage[varg]{txfonts}   
\usepackage{subfigure} 

\def\ga{\mathrel{\mathchoice {\vcenter{\offinterlineskip\halign{\hfil
$\displaystyle##$\hfil\cr>\cr\sim\cr}}}
{\vcenter{\offinterlineskip\halign{\hfil$\textstyle##$\hfil\cr>\cr\sim\cr}}}
{\vcenter{\offinterlineskip\halign{\hfil$\scriptstyle##$\hfil\cr>\cr\sim\cr}}}
{\vcenter{\offinterlineskip\halign{\hfil$\scriptscriptstyle##$\hfil\cr>\cr
\sim\cr}}}}}

\wocname{ISVHECRI}
\woctitle{ISVHECRI 2012}
\begin{document}
\title{Experimental Summary:\\ Very High Energy Cosmic
Rays and their Interactions}
%
%

\author{Karl-Heinz Kampert\inst{1}\fnsep\thanks{\email{kampert@uni-wuppertal.de}}}

\institute{University Wuppertal, Department of Physics}

\abstract{%
The XVII International Symposium on Very High Energy Cosmic Ray Interactions, held in August of 2012 in Berlin, was the first one in the history of the Symposium, where a plethora of high precision LHC data with relevance for cosmic ray physics was presented. This report aims at giving a brief summary of those measurements and it discusses their relevance for observations of high energy cosmic rays. Enormous progress has been made also in air shower observations and in direct measurements of cosmic rays, exhibiting many more structure in the cosmic ray energy spectrum than just a simple power law with a knee and an ankle. At the highest energy, the flux suppression may not be dominated by the GZK-effect but by the limiting energy of a nearby source or source population. New projects and application of new technologies promise further advances also in the near future. We shall discuss the experimental and theoretical progress in the field and its prospects for coming years.
}
\maketitle
\section{Introduction}
\label{intro}
The Seventeenth International Symposium on Very High Energy Cosmic Ray Interactions has brought together high energy physicists from the domains of collider and fixed target experiments and from cosmic ray observatories. An impressive set of data has been collected since the previous ISVHECRI in both domains and has advanced the field a lot. Accelerator experiments have started to give crucial information for the modelling of high energy hadronic interactions relevant for air shower simulations and have helped to reduce systematic uncertainties in reconstructing primary energy and mass from extensive air shower experiments. At the same time, cosmic ray experimenters have improved their measurements enabling them to see much more structure in the energy spectrum than just a simple power law distribution. These data come along with much better composition information so that the underlying astrophysics is starting to be decomposed.

\begin{figure}
\centering
\subfigure{\includegraphics[width=0.8\columnwidth]{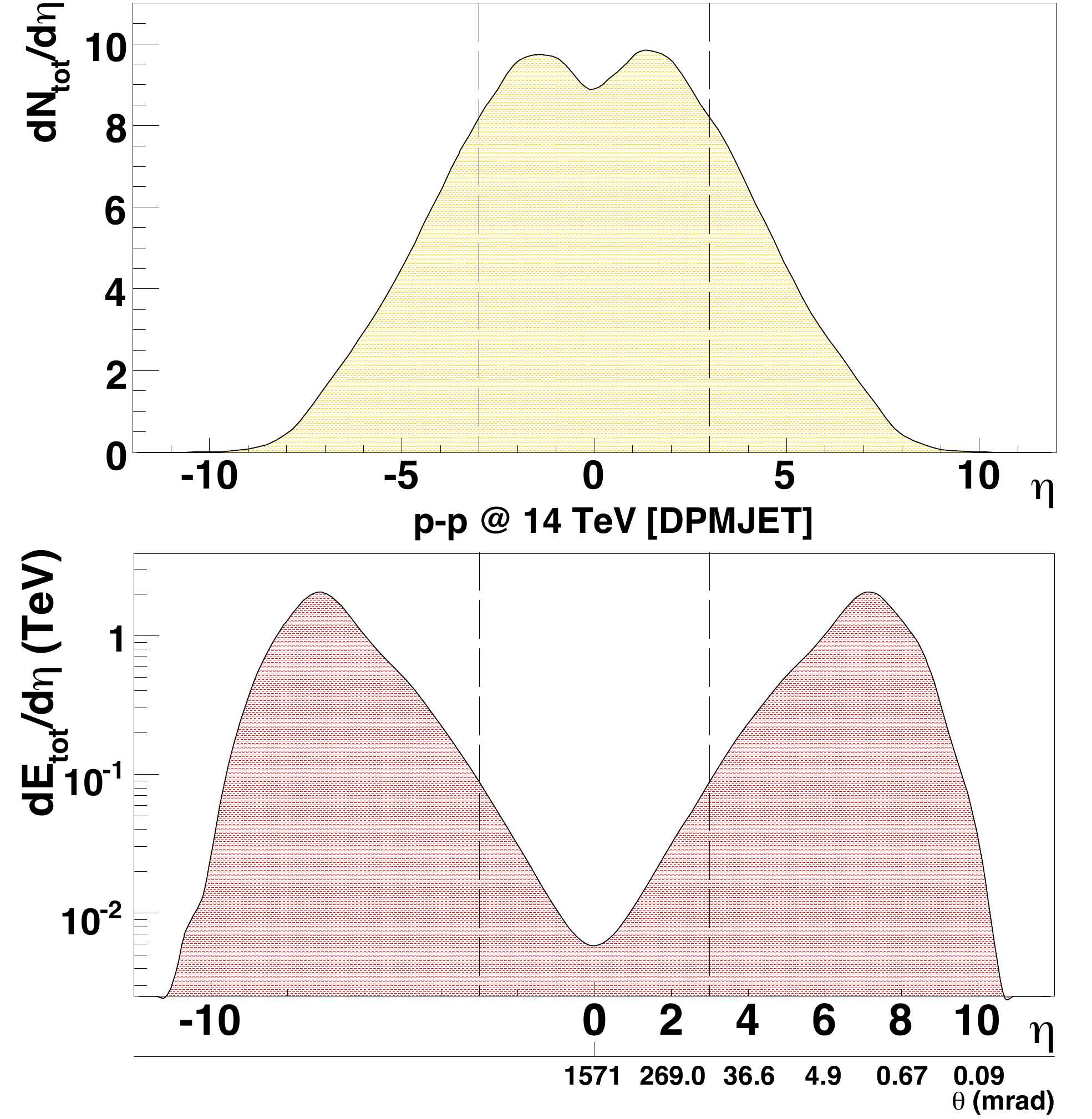}}
\subfigure{\includegraphics[width=0.8\columnwidth]{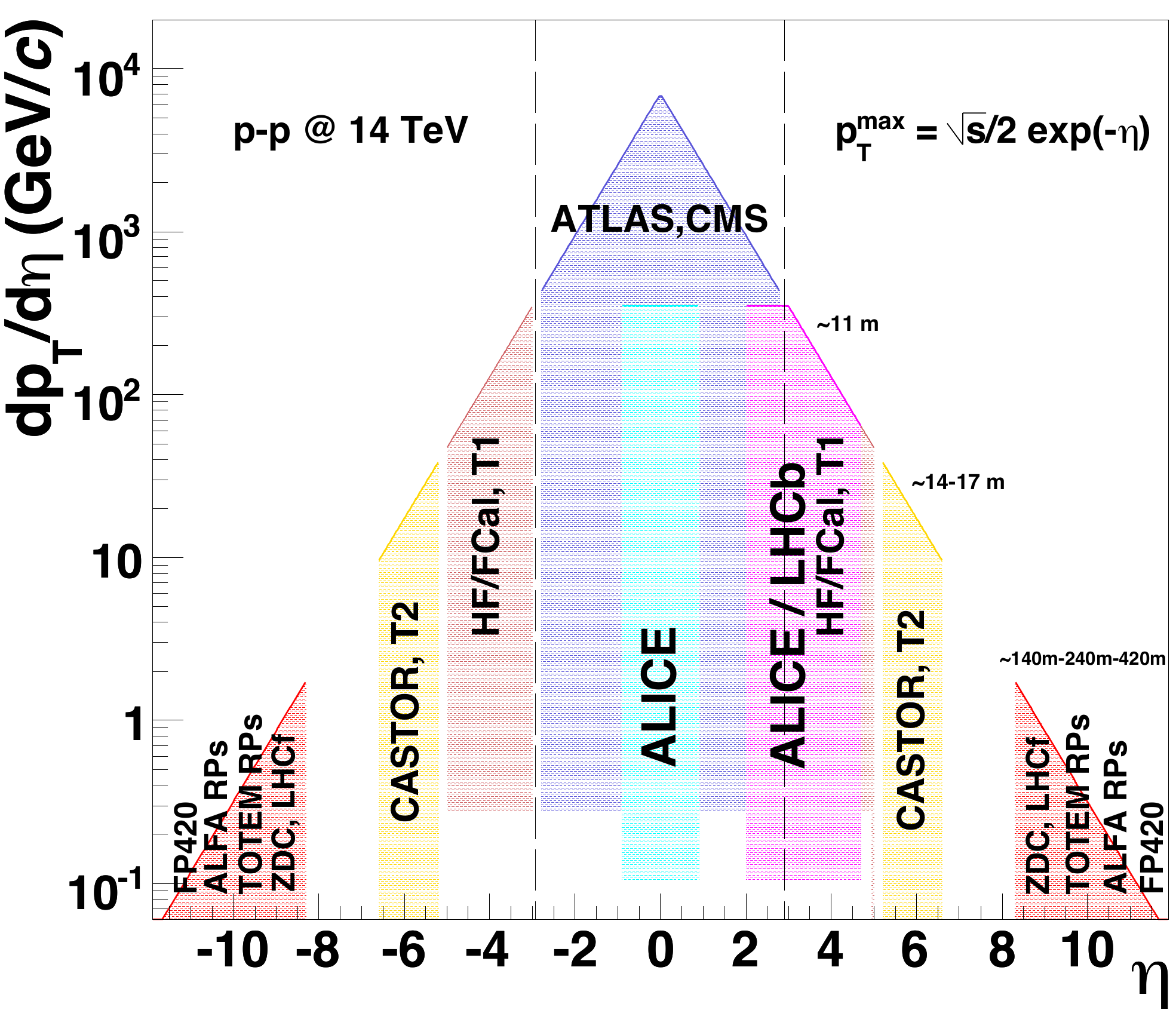}}
\caption{Sketch of the particle and energy flow in pp collisions at $\sqrt{s}=14$\,TeV and compilation of phase space acceptances by various LHC experiments \cite{Enterria-08}.}
\label{fig:acceptances}
\end{figure}

It is the first meeting in this series, where a plethora of new precision data from LHC has become available up to $\sqrt{s}=8$\,TeV and for p+p, p+Pb, and Pb+Pb collisions. The discovery of a Higgs-like particle by CMS and ATLAS has been celebrated a lot and was reported extensively in public media. Congratulations to the two collaborations for this great achievement, also reported about at this meeting. However, low cross section processes, such as the Higgs or top-quark production are of secondary importance for cosmic ray interactions. Instead, the bulk processes observed in minimum bias events with about $10^{10}$ times the Higgs production cross sections are of prime interest, particularly those leading to very forward (large rapidity) particle production. This is because, different from the particle flow which peaks at mid-rapidity, most of the beam energy actually is observed close to projectile (and target) rapidities and remains mostly unobserved in the beam-pipes. This region of phase space is the one that drives the evolution of extensive air showers. It is thus of prime importance to extend the acceptance with dedicated detector subcomponents to as close to the beam pipe as possible. Figure\,\ref{fig:acceptances} presents a compilation of the phase space acceptances reached in current experiments at LHC.

Besides collider data from LHC and still from HERA, fixed target experiments continue to give valuable information about hadronic collisions in an energy domain where the majority of hadronic interaction take place during the development of air showers. Close cooperation and dialog between the communities is now especially timely, and it is gratifying to see the symposium equally well attended by high-energy physicists and model builders.

In this summary, we shall start with discussing recent data from accelerator experiments and with their implications for simulations of extensive air showers (EAS), address the observed deficit of muons in EAS simulations, discuss news from emulsion experiments and then move on to latest results from air shower experiments and high energy neutrino detectors.

\section{Accelerator Experiments}
\label{sec:accelerator}
\subsection{Particle and Energy Flow}
\label{sec:particle_flow}

Data from accelerator experiments have guided the development of hadronic interaction models from the early beginnings. A major step forward has now been taken with the advent of LHC data. Properties of interest for the modeling of EAS are the overall particle and energy flow as a function of rapidity, which in absence of full particle identification is often approximated by the pseudorapidity $\eta = -\ln (\tan \theta/2)$, shown in Fig.\,\ref{fig:acceptances}. The transverse momentum, $d\sigma/dp_\perp$, particle multiplicity distribution, $dN/dM$ (with $M$ being the charged particle multiplicity), as well as the total-, inelastic-, and diffractive cross sections, $\sigma_{\rm tot}$, $\sigma_{\rm inel}$, $\sigma_{\rm diffr}$, respectively, are further observables of prime interest.

\begin{figure}
\centering
\sidecaption
\includegraphics[width=\columnwidth]{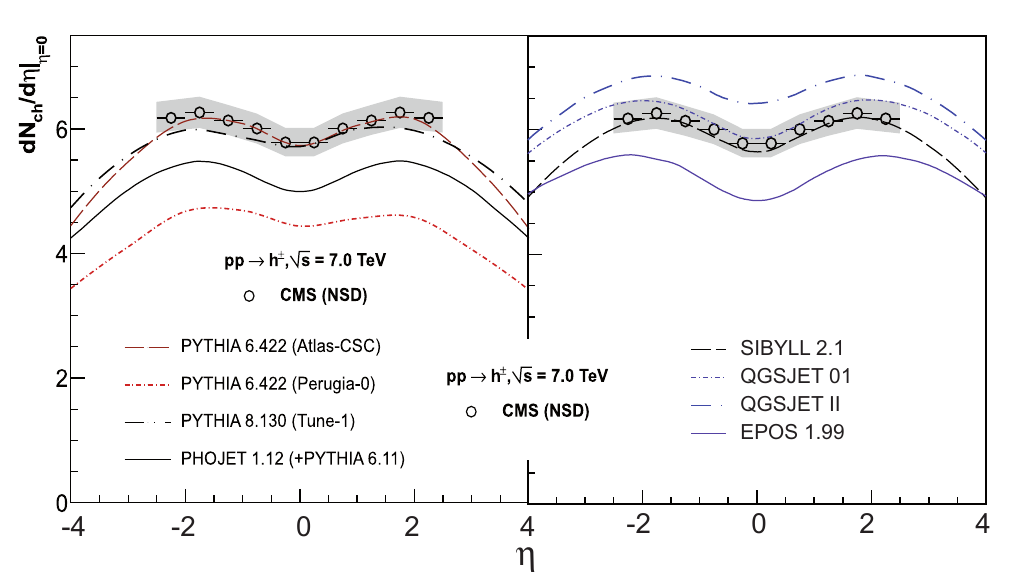}
\caption{Pseudorapidity distributions of charged hadrons in NSD pp interactions measured by CMS at $\sqrt{s}=7$\,TeV. The grey band indicates the systematic uncertainties. The data are compared to typical HEP and EAS models. (Adapted from \cite{Enterria-11}.)}
\label{fig:dndeta}
\end{figure}

$dN_{\rm ch}/d\eta$ distributions for non-single diffractive (NSD) pp-interactions at $\sqrt{s}=7$~TeV are shown in Fig.\,\ref{fig:dndeta} and are compared to the most common HEP models PYTHIA and PHOJET (left) and to interaction models common in EAS modeling (right). While PYTHIA and PHOJET underestimate the pseudorapidity distributions by up to a significant amount, models applied in EAS simulations bracket the experimental data. Moreover, it should be stressed, that no extra tuning has been done in the latter case, i.e.\ the simulated curves are predictions \cite{Enterria-11}. This comparison gives some confidence in the EAS models and also supports the general practice that the interpretation of EAS data with different EAS interaction models provides an estimate of the systematic uncertainties involved. However, looking at the energy dependence, we note that EPOS 1.99 grows too slowly and QGSJET II too fast with increasing cms energy. Thus, data at $\sqrt{s}=14$\,TeV, hopefully available after the long 2013/2014 shut down, will be of great interest. As shown by A.\,de\,Roeck \cite{this-volume}, $dN/d\eta|_{\eta=0}$ of NSD events predicted by EPOS 1.99 and QGSJET II exhibits differences by a factor 2-3 when extrapolating into the GZK domain at $\sqrt{s}\simeq 400$\,TeV.

\begin{figure}
\centering
\includegraphics[width=0.9\columnwidth]{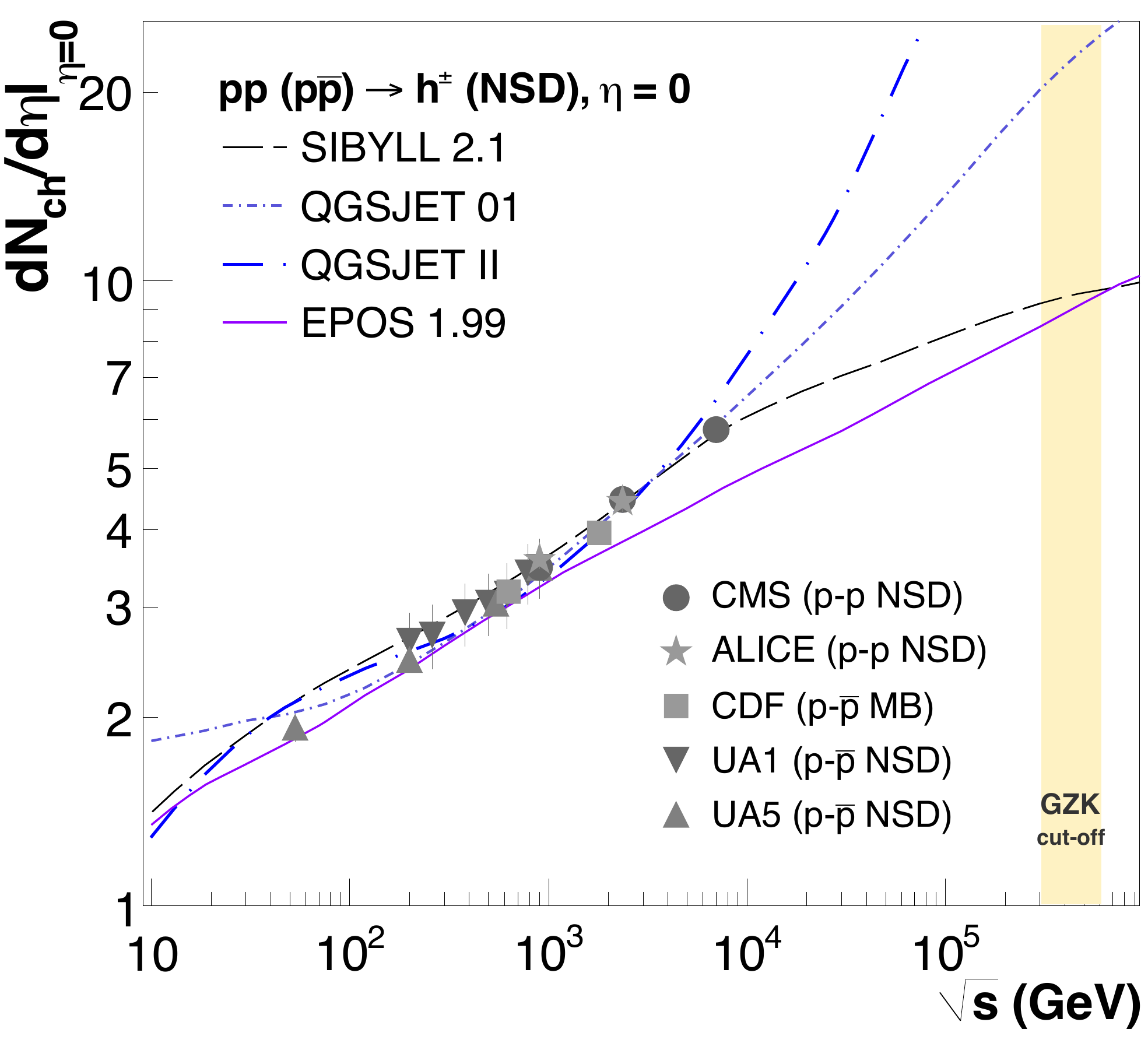}
\caption{Evolution of  $dN/d\eta|_{\eta=0}$ for NSD events as a function of the cms energy. Large differences are seen for extrapolations into the GZK domain at $\sqrt{s}\simeq400$\,TeV   \cite{Enterria-11}.}
\label{fig:multiplicty}
\end{figure}

A very good description of the energy flow $dE/d\eta$ measured by CMS in Pb+Pb collisions up to $\eta=6$ and in different bins of centrality is provided by EPOS 1.99 and QGSJET II (see presentations by I.\,Katkov, D.\,Volanskyy and de\,Roeck \cite{this-volume}). However, EPOS and SIBYLL fail again in describing the very forward ($9.2 < \eta < 9.4$) $\pi^0$ production measured by LHCf at large transverse momentum of $p_\perp \ga 0.4$\,GeV/c (see presentation by Y.\,Itow \cite{this-volume}). Similarly, H1 data from HERA are still being analyzed and the photon production (dominated by $\pi^0$-decay) up to large values of $x_F$ is again well described by QGSJET, but only poorly by EPOS and Sibyll (see A.\,Buniatyan at this conference \cite{this-volume}). These new results - even though still bracketed by the models - indicate that EPOS and SIBYLL produce too hard $\pi^0$ spectra at forward rapidities. 

At this stage we may conclude that the bulk data on energy and particle flow are fairly well described by EAS models but more improvements are needed.  This requires joining forces between the CR and HEP community. Model builders of the CR-community need to identify which data are most important for further tuning of their models and they also need to figure out, what level of precision would be needed for different observables. Only with such kind of information and prioritization, can additional measurements can be planned and detectors appropriately designed. A very useful study in this direction was recently performed by Ulrich, Engel and Unger \cite{Ulrich-11}. It demonstrated that the interaction cross section, charged particle multiplicity, elasticity, and charge ratio influence the EAS observables ($N_e$, $N_\mu$, $X_{\rm max}$, and RMS $X_{\rm max}$) in very specific correlated and anti-correlated ways.

\subsection{Observation of Muon Bundles with Accelerator Experiments}
\label{sec:muon-bundles}
Observations of muon bundles with very large muon multiplicities were first reported by LEP experiments (see e.g.\ \cite{Avati-03,Delphi-07}). These events were observed at a rate almost an order of magnitude larger than expected from simulations even when making extreme assumptions on the measured all-particle flux and assuming a pure iron composition and as such often were interpreted in terms of new physics.

\begin{figure}
\centering
\sidecaption
\includegraphics[width=0.8\columnwidth]{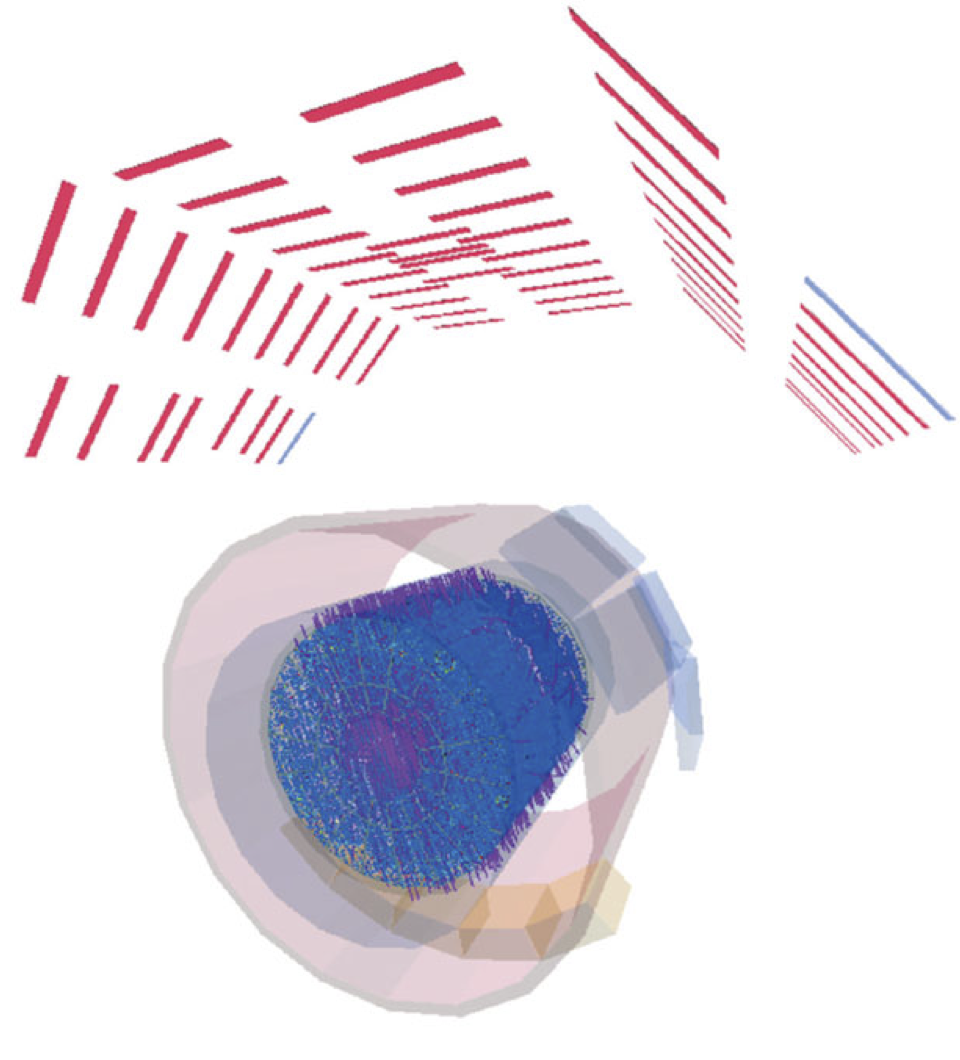}
\caption{Example of a high-multiplicity cosmic ray event in ALICE which fired most scintillator trigger modules (ACORDE) and yields a density of $\sim 18$ muons/m$^2$ within the TPC volume \cite[Grosse-Oetringhaus]{this-volume}.}
\label{fig:muon-bundle}
\end{figure}

An update on such measurements was presented by J.\,F.\,Grosse-Oetringhaus reporting data from the ALICE TPC \cite{this-volume}. Again, muon densities up to 18 m$^{-2}$ were found within the TPC volume yielding more than 270 muons observed in single events during 11 days of effective data taking in 2010/2011. Careful modeling with CORSIKA using the all-particle spectrum as input and assuming an increasingly heavy mass composition beyond the knee (see below) recently demonstrated that such events are expected at the measured rate from iron dominated EAS reaching energies near to $10^{17}$\,eV and landing some tens of metres from the ALICE detector \cite{Alessandro-13}. This still preliminary finding appears to eliminate the need of exotic physics and supports measurements of a very heavy composition beyond the knee. However, as shall be discussed below, EAS observations at the highest energies still show a deficit of muons in EAS models which may reach a factor of 2 for a proton dominated composition and is reduced to about 30\,\% when assuming a iron dominated composition.

\subsection{Measurements of Cross Sections}
\label{sec:X-sections}

A very nice introduction to cross-section measurements was given by J.\,Pinfold \cite{this-volume}. Generally, the total cross section is disentangled into different components
\begin{equation}
\sigma_{\rm tot} = \sigma_{\rm elastic} + \sigma_{\rm inelastic}
\end{equation}
with:
\begin{equation}
\sigma_{\rm inelastic} = \sigma_{\rm non-diff} + \sigma_{\rm single-diff} + \sigma_{\rm double-diff} + \sigma_{\rm excl} + \ldots
\end{equation}

It is probably fair to say that the TOTEM experiment, presented at this symposium by K.\,Eggert \cite{this-volume}, performed the most advanced and precise measurements of the pp elastic, -total, -inelastic, and -diffractive cross sections at LHC energies. It employs two telescopes T1 and T2 at $3.1 < \eta < 4.7$ and $5.3 < \eta < 6.5$, respectively, for the measurements of the inelastic cross section and 24 Roman pots up to a distance of 220\,m on both sides of the interaction point for measurements of the elastic and diffractive pp cross-section. From a very detailed analysis of the diffractive pattern and extrapolation to $|t|=0$ with $t=-p^2 \cdot \theta^2$ the integrated elastic cross section is found to be $\sigma_{\rm elast} = 25.4 \pm 1.0^{\rm lumi} \pm 0.3^{\rm syst} \pm 0.03^{\rm stat}$\,mb \cite{Eggert}. For the inelastic cross section $\sigma_{\rm inel} = 73.7 \pm 0.1^{\rm stat} \pm 1.7^{\rm syst} \pm 2.9^{\rm lumi}$\,mb and for the total cross section then $\sigma_{\rm tot} = (98.3 \pm 2.0)$\,mb is reported \cite{Totem-11}. Consistent results were reported by Pinfold,
de\,Roeck, and Grosse-Oetringhaus \cite{this-volume}.

\begin{figure}
\centering
\sidecaption
\includegraphics[width=\columnwidth]{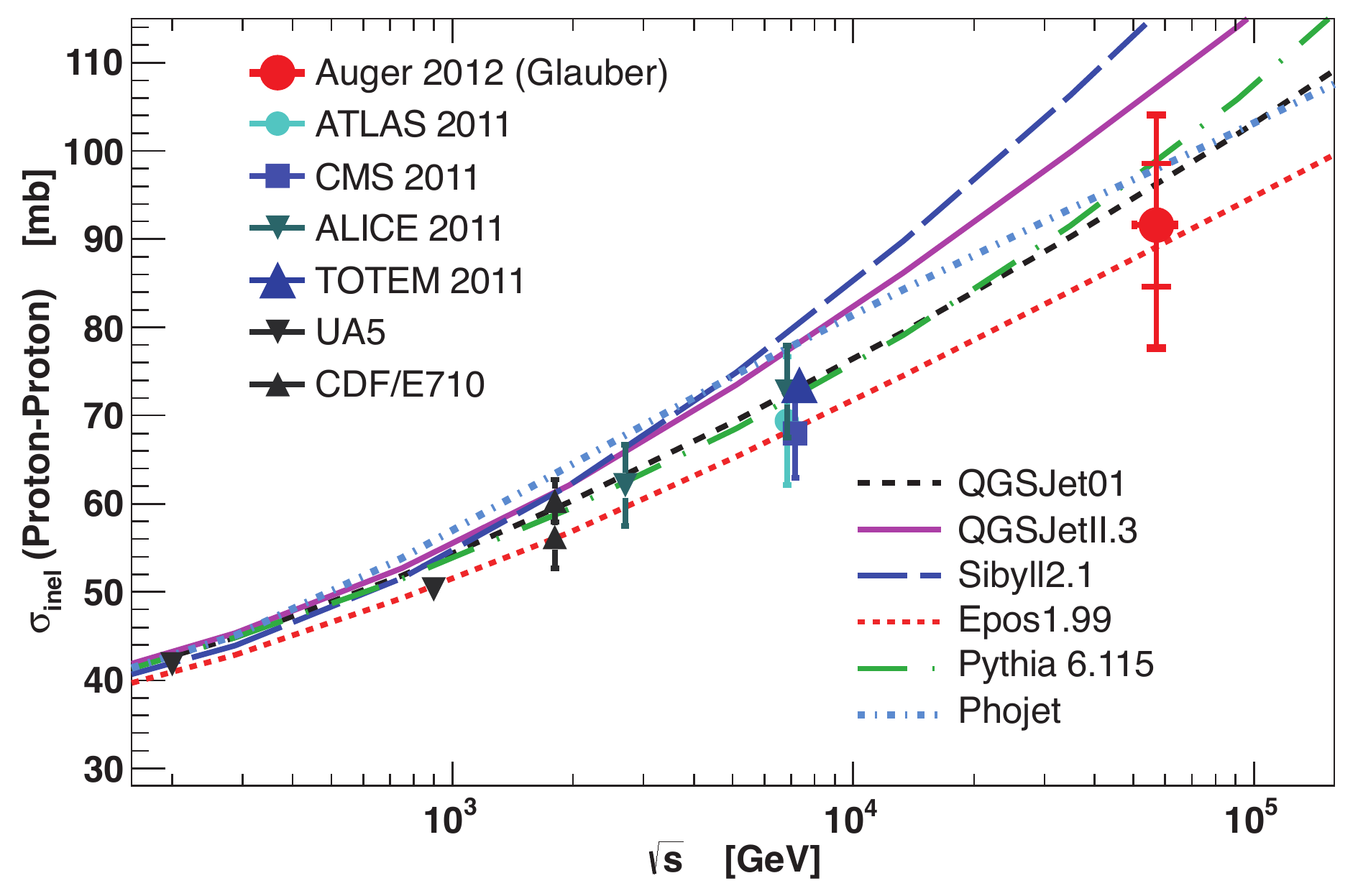}
\caption{Comparison of derived $\sigma_{pp}^{\rm inel}$ from Auger to model predictions and accelerator data, discussed at this conference. The inner error bars of the Auger data represent statistical errors and the outer ones include systematic uncertainties \cite{Auger-xsection-12}.}
\label{fig:x-section}
\end{figure}

Very interesting measurements of the pp inelastic and total cross sections at $\sqrt{s}=57$\,TeV, i.e.\ almost ten times the LHC energy,  were reported by R.\,Ulrich based on EAS data from the Pierre Auger Observatory \cite{this-volume}. This is done by selecting CR primaries at $10^{18} < E/{\rm eV} < 10^{18.5}$ and analyzing the exponential tail of the $X_{\rm max}$ distribution $dn/dX_{\rm max} \propto \exp{\left(-X_{\rm max}/\Lambda\right)}$ for a total of about 800 events passing all cuts. The attenuation parameter $\Lambda$ is then inversely related the p-air total cross section. In order to properly account for effects, such as of shower-by-shower fluctuations related to the point of first interaction in the atmosphere, the cross section has been inferred from EAS simulations with slightly modified cross sections. To convert from p-air to pp cross-section, a modified Glauber approach has been performed with the result shown in Fig.\,\ref{fig:x-section} together with the aforementioned LHC results and those from the Tevatron and CERN SPS.

\subsection{Fixed Target Experiments}
\label{sec:fixed-target}

\begin{figure}
\centering
\sidecaption
\includegraphics[width=\columnwidth]{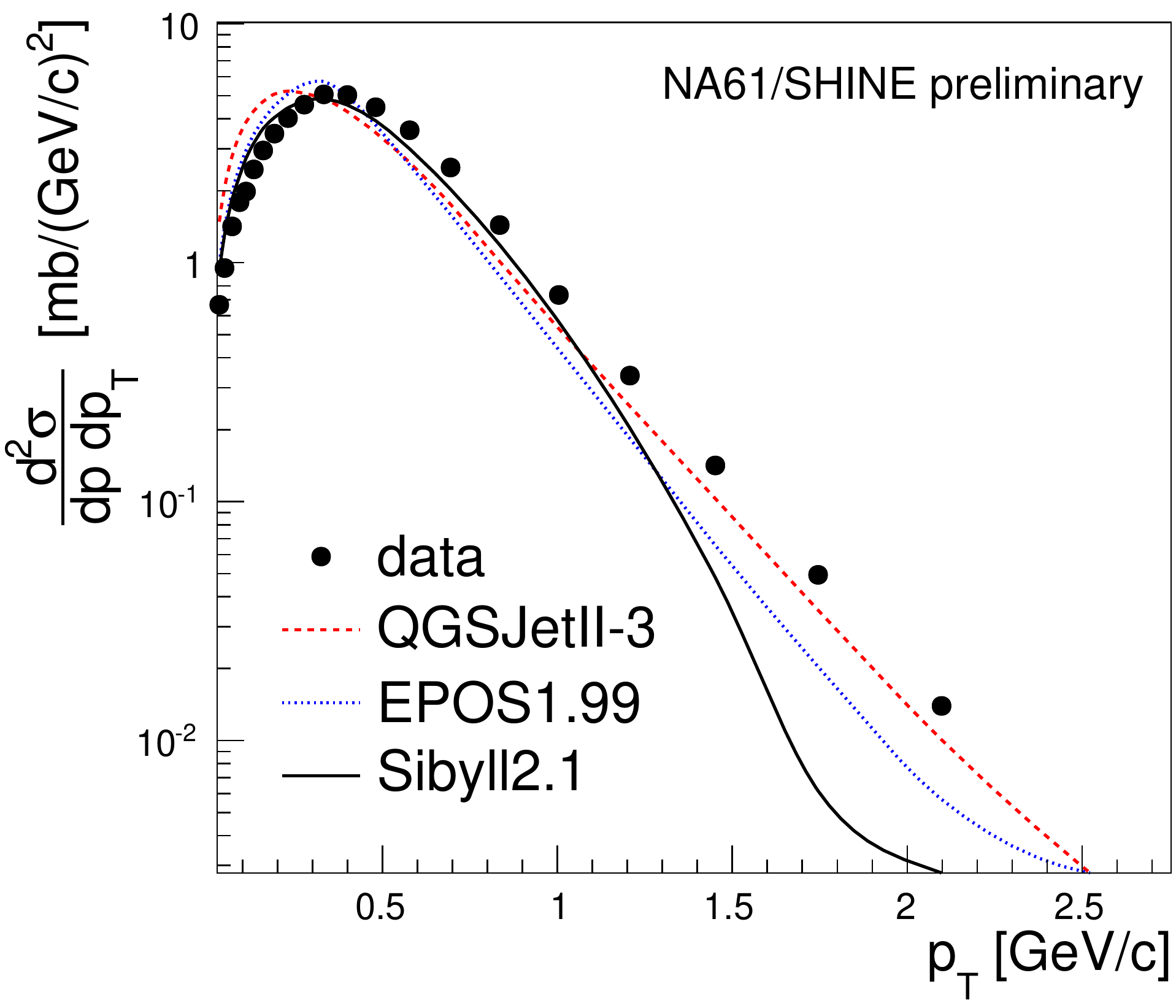}
\caption[CR data MC comparison]{Transverse momentum spectrum of
negatively charged hadrons produced in $\pi^-$+C interactions at
158~GeV/$c$ beam momentum at $\langle p\rangle=$10.4~GeV/$c$ \cite[Unger]{this-volume}.}
\label{fig:NA61}
\end{figure}

The fixed target program has almost vanished at CERN because of interests focussed strongly to the highest cms-energies and to tests of the standard model of particle physics. However, NA61/SHINE is taking data at the SPS with its science program -- besides searching for the onset of deconfinement and for the critical point of strongly interacting matter and providing valuable data for neutrino experiments  -- being well aligned with the needs of the cosmic ray community. There is a full program still going on for another few years measuring a wide range of projectile-target combinations across a range of beam energies from 13 - 160 GeV/nucleon. Its also worthwhile mentioning that the collaboration involves scientists from nuclear, particle, and cosmic ray physics. A report about recent measurements was given by M.\,Unger and M.\,Markariev \cite{this-volume}.
Figure \ref{fig:NA61} shows an exemplary plot of negatively charged hadron production in $\pi^-$+C interactions at 158\,GeV/$c$ beam momentum at $\langle p \rangle = 10.4$\,GeV/$c$. The data are compared to recent versions of high energy hadronic interaction models employed in CORSIKA. Here, EPOS\,1.99 and QGSJet\,II-3 show a very good agreement while Sibyll\,2.1 fails desribing the high $p_\perp$-part of the spectrum.

Comparisons of $\pi^+$ production data in p+C reactions at 31\,GeV/$c$ to the two commonly used low-energy interaction models GHEISHA and Fluka exhibited severe problems in GHEISHA \cite[Unger]{this-volume}. This model has served for a long time e.g.\ in detector simulations with GEANT.

The statistics of the p+C data will be increased by another factor of 10. Moreover, baryon and anti-baryon, strange baryons, and $\rho^0$-mesons will be measured in the near future, all of which will be of utmost importance for cosmic ray measurements, particularly in view of trying to solve the deficit of muons observed in comparison to EAS data (see next section).


\section{The Muon Problem in EAS-Simulations}
\label{sec:mu-problem}

\begin{figure}
\centering
\sidecaption
\includegraphics[width=\columnwidth]{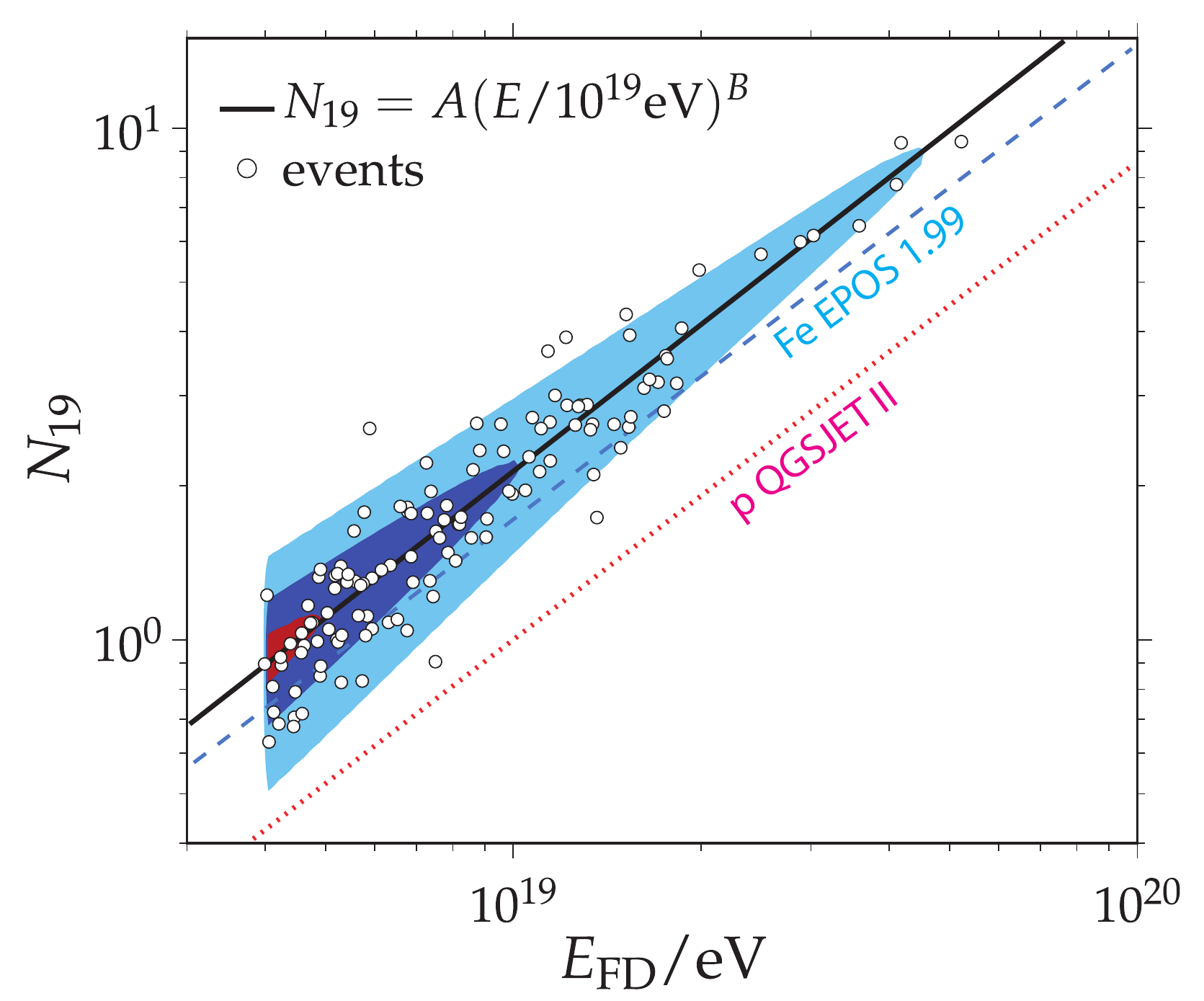}
\caption{Fit of a calibration curve $N_{19} = A(E/10~{\rm EeV})^B$. The constants $A$ and $B$ are obtained using the maximum-likelihood method. The contours indicate the constant levels of the p.d.f.\ integrated over zenith angle, corresponding to 10, 50 and 90\% of the maximum value~\cite[H.\ Dembinski]{ICRC-Comp-I}. Calibration curves for protons QGSJETII (dot line) and iron EPOS1.99 (dashed line) are shown for comparison.}
\label{fig:muons-auger}
\end{figure}

The so called muon-problem in EAS-simulations dates back more than a decade by now. It was reported first by the prototype HiRes telescope when operated at the MIA air shower array \cite{HiRes-00}. When comparing the muon density measured by MIA as a function of CR-energy determined in a calorimetric way by the HiRes-Telescope, a deficit of muons in the simulations was found with respect to MIA data when a composition lighter than Fe was assumed. A proton dominated composition, however, was preferred from the $X_{\rm max}$ measurement of the same HiRes prototype telescope. Similar observations were later  reported by the multi-detector array KASCADE and were reported for the successor experiment KASCADE-Grande at this meeting by A.\,Haungs \cite{this-volume}. He showed that the muon deficit around $10^{17}$\,eV can almost be neglected within the systematic uncertainties for vertical showers but increases significantly for larger zenith angles up to $40^\circ$. Going to higher energies, P.\,Sokolsky in his review pointed out that there is a 27\,\% difference in the reconstructed energy of the Telescope Array (TA) dependent whether the calibration is based in the fluorescence telescopes or on simulations of the ground array \cite{this-volume}. Very similarly, J.\,Allen showed the same effect increasing again for larger zenith angles \cite{this-volume}. This is shown in Fig.\,\ref{fig:muons-auger} for events at $\theta > 60^\circ$ and energies ranging from $4\cdot 10^{18}$\,eV to $6\cdot 10^{19}$\,eV. Here, $N_{19}$ is defined as the ratio of the total number of muons, $N_{\mu}$, in the shower with respect to  the total number of muons at $E=10$\,EeV given by a 2-dim reference distribution, $N_{19}=N_{\mu}(E,\theta) / N^{{\rm map}}_{\mu}(E=10~{\rm EeV},\theta)$ which accounts for the geomagnetic spatial deviation of muons at ground ~\cite[Dembinski]{ICRC-Comp-I}. As shown in the figure, the muon number estimated from QGSJET-II is lower by almost a factor of two compared to data and a pure iron composition simulated in EPOS\,1.99 -- which yields the highest muon numbers of all simulations done -- still is too low by about 20\,\% compared to data. 

Thus, all of the experiments and all of the different analyses show a significant deficit in the number of muons predicted by simulations. This discrepancy cannot be explained by the mass composition alone, although a heavy composition reduces the relative excess of the data down to $N_\mu^{\rm data}/N_\mu^{\rm model} \sim 1.3$. Moreover, the observed zenith angle dependence of $N_{\mu}^{\text{rel}}$ suggests that, in addition to the number, there may also be a discrepancy in the attenuation and lateral distribution of muons between the simulations and data \cite{Kampert-ICRC11}.

The origin of this deficit is not yet understood and possible reasons were discussed by S.\,Ostapchenko, L.\,Cazon, R.\,Conceicao and others (see theoretical summary by S.\,Sarkar) \cite{this-volume}. Those include missing $\rho^0$ production, too low (anti-)baryon and/or kaon production, a suppression of $\pi^0$ production due to chiral symmetry restoration above some threshold energy, etc.

\section{Direct CR-Observations up to the Knee}
\label{sec:direct}
\subsection{Recent Results from Emulsion Experiments}
\label{sec:emulsions}

Emulsions, due to their excellent spatial resolution and tracking capabilities along with information about the specific energy losses of individual particles, look back to a long and successful history in cosmic ray and particle physics. However, when looking for rare events and trying to quantify their abundance, not only the average tracks need to be understood, but also their fluctuations on an event-by-event basis. Thus, the tool to be brought into place is detailed Monte Carlo simulations of the emulsion response to a flux of cosmic ray particles. This major step is now beginning to be taken and results were presented by A.S.\,Borisov and M.\,Tamada \cite{this-volume}.

\begin{figure}
\centering
\sidecaption
\includegraphics[width=\columnwidth]{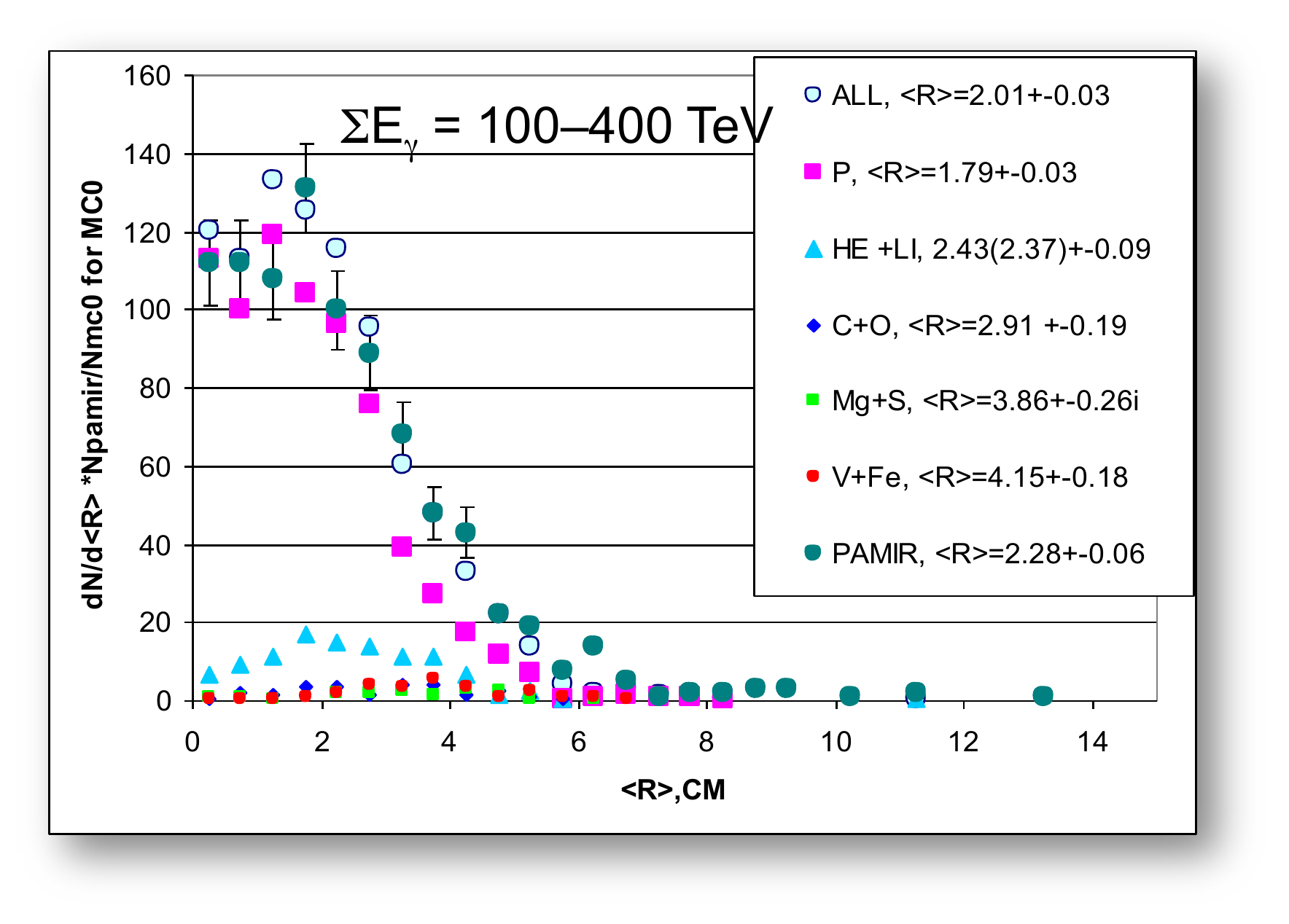}
\caption{Radial extension of $\gamma$-families found in emulsions for data and different primary masses simulated and measured for $100 < \sum E_\gamma/TeV < 400$~\cite[Borisov]{this-volume}.}
\label{fig:emulsions}
\end{figure}

Borisov presented such studies for the radial extension of $\gamma$-families binned into intervals of total energy deposited in the emulsions (Fig.\,\ref{fig:emulsions}) \cite{this-volume}. As demonstrated by the figure, the observed radial spread $\langle R_{\rm data} \rangle = (2.28\pm0.06)$\,cm can be described by a proton and He dominated Galactic cosmic ray composition (open circles), with $\langle R_{\rm p} \rangle = (1.79\pm0.063)$\,cm and $\langle R_{\rm He} \rangle = (2.43\pm0.093)$\,cm. Another interesting development was presented by Tamada and Borisov where a hybrid arrangement of an air shower array, burst detector, and $\gamma$-family detector is operated. Thus, the correlations amongst the observables can be studied and compared with CORSIKA+GEANT simulations. While the energy deposited in emulsions, $\sum E_\gamma$, shows the expected correlation with the electron shower size, $N_e$, of the array, deviations between data and simulations are found in the correlation of $\sum E_\gamma$ with the number of detected bursts, $n_b$. Whether this would point to new physics or to some yet unknown detector or simulation effect remains to be solved. 

Very ambitious plans for a more complex hybrid experiment, named PAMIR XXI, were presented by Borisov \cite{this-volume}. The goal of the work done within the Pamir-Chacaltaya ISRC at a height of  4300\,m a.s.l.\ is to study the event properties and cosmic ray composition up to the knee energy. For this purpose, the air shower array is planned to cover an area of $80 \times 80$\,m$^2$. It will be complemented by a large calorimeter, open Cherenkov detectors, and imaging Cherenkov telescopes. More details will be worked out in a dedicated workshop planned in the near future at the Pamirs.

To summarize this section, I think the advent of hybrid observations between electronic and emulsion detectors in cosmic ray physics is very gratifying to see. A successful combination of such detectors is already known e.g.\ from the OPERA neutrino experiment \cite{OPERA} at Gran Sasso laboratory.  However, before claiming observation of new physics in any experiment or detector, information about the number of observed events and integrated exposure would be required. Moreover, full distributions of experimental observables are required after application of acceptance cuts for measured {\em and} for (e.g.\ CORSIKA + GEANT) simulated events so that the overall agreement of data and Monte Carlo can be verified and the fraction of observed events showing exotic features be quantified. This is a necessary step also for estimations of the cross section of such processes. Clearly, the first steps have been taken and this method deserves more thorough application.


\begin{figure}
\centering
\sidecaption
\includegraphics[width=\columnwidth]{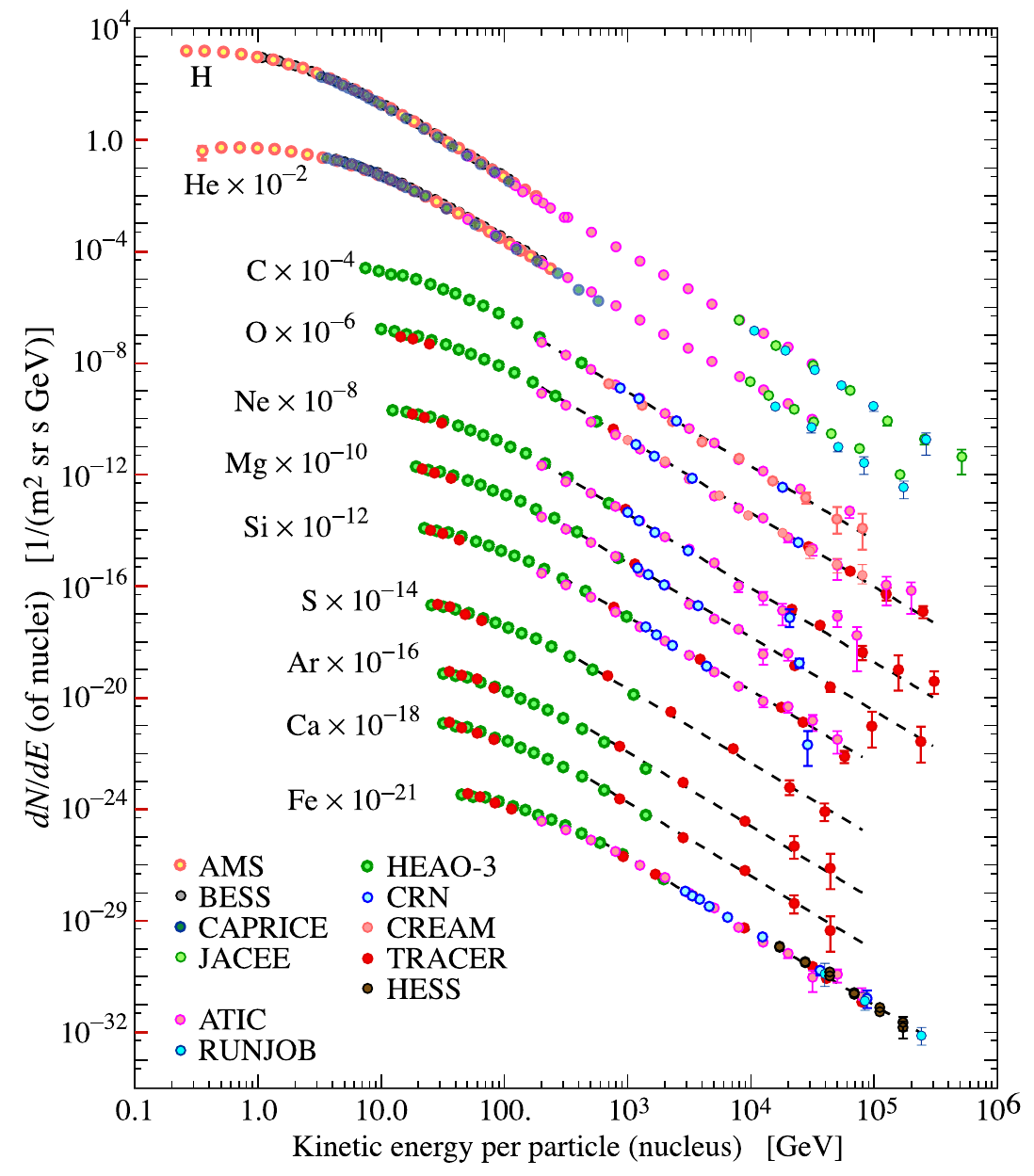}
\caption{Compilation of energy spectra from direct experiments \cite{PDG}.}
\label{fig:spectra}
\end{figure}

\subsection{Balloon and Satellite Observations}
\label{sec:satellite}
A comprehensive review of direct measurements of cosmic rays was presented by R.\,Sparvoli \cite{this-volume}. Clearly, data from recent experiments have reached a new level of quality and quantity, with PAMELA taking the lead. The energy spectra of protons and nuclei start to exhibit structures beyond simple power laws and also show different shapes for different nuclei (c.f.\,Fig.\,\ref{fig:spectra}). Perhaps most importantly, the hardening of He-spectra relative to protons at rigidities above some $\sim 10$\,GV -- discussed also by Di Sciascio from CREAM \cite{this-volume} -- seems to continue for heavier nuclei. As a consequence, the He/p-ratio increases with possibly the overall mass composition at the knee becoming significantly heavier than the standard Galactic cosmic ray composition. This expectation is supported also by recent EAS data (see next section). PAMELA also reported an abrupt hardening of the p- and He-spectra at $R\simeq235$\,GV with a significance of 95\,\% CL. Since propagation effects should be negligible above 100\,GV, these findings - if confirmed - may provide clues about the underlying source spectra.

Antiparticles are another interesting topic because they may provide information about annihilation of dark matter particles and also constitute stringent tests about Galactic propagation. Recent PAMELA data confirm the observed rise of the positron fraction $\phi(e^+)/(\phi(e^+) + \phi(e^-))$ to energies up to about 200\,GeV and are not described easily by Galactic propagation. Most recently, the AMS Collaboration has released its first data with unprecedented statistical accuracy \cite{AMS-positrons} which confirm the general trend and agree very well with PAMELA. Antiprotons, measured by PAMELA, BESS and others do - on the other hand - not show any significant deviation from expectations based on Galactic propagation, i.e.\ they are well described as being produced by CR interactions with the interstellar medium. This principal difference between positrons and antiprotons is a challenging puzzle to be solved. Besides being interpreted as DM annihilation, a more conventional  astrophysical interpretation of the positron excess may by an emission from mature pulsars ($\cal{O}$$(10^5)$\,years) \cite{Hooper-09}. The continuation of the positron spectrum towards higher energies should settle this question in the near future after the advent of AMS data. Moreover, the signal from nearby pulsars is expected to generate a small but significant dipole anisotropy in the cosmic ray electron spectrum, potentially providing a method by which gamma-ray space telescopes would be capable of discriminating between the pulsar and dark matter origins of the observed high energy positrons. With no anisotropies having been found yet, the question about the origin of the positron excess remains to be answered.

\begin{figure}
\centering
\sidecaption
\includegraphics[width=\columnwidth]{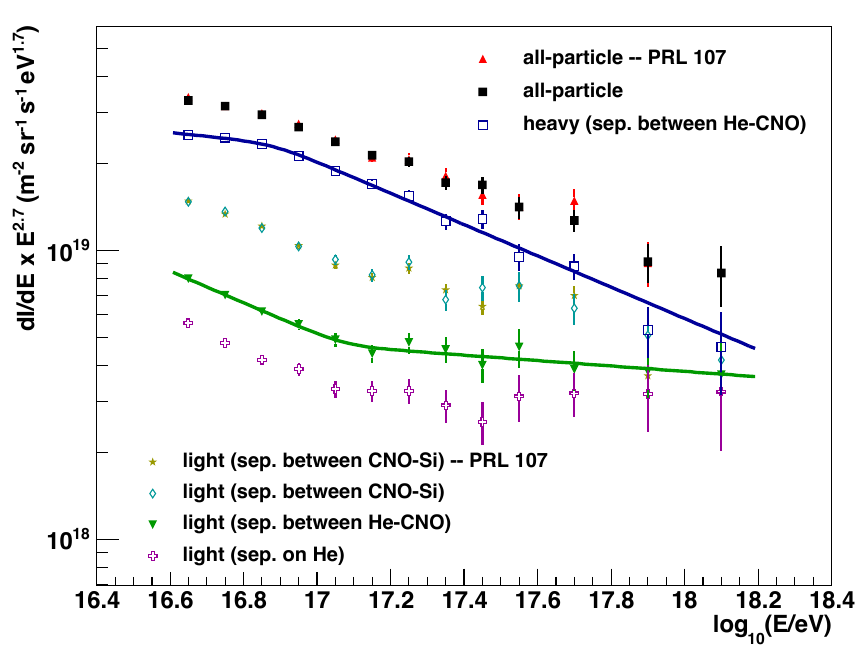}
\caption{The all-particle (full black squares) and electron-poor (=heavy) spectra (open squares) from \cite{KG-Fe-knee} in comparison to electron-rich (=light) spectra (triangles) \cite{p-ankle}. The lines represent broken power law fits to the data.}
\label{fig:spectra1}
\end{figure}

\section{Results from Air Shower Experiments}
\label{sec:eas-expts}
\subsection{Two Knees in the CR-Spectrum}
\label{sec:knee}

A.\,Chiavassa presented an overview of the latest results in the knee energy range \cite{this-volume}. As he pointed out, the knee is a well established phenomenon of yet unknown astrophysical origin rather than a feature of changing hadronic interactions. Measurements based on electron and muon numbers, performed e.g.\ by EAS-TOP and KASCADE, have demonstrated an increasingly heavy composition above the knee. Furthermore, a 2d-unfolding of the KASCADE $N_e$ vs $N_\mu$ numbers suggests a dominance of p and He-nuclei at the knee and an energy shift of the break in the energy spectra of different mass groups which is compatible with the assumption of a constant rigidity. The strong abundance of He at the knee has been a surprise, but this finding together with the increasingly heavier composition above the knee is now well established and confirmed by other air shower experiments (see e.g.\ D.\,Di Sciascio's report from ARGO-YBJ or S.\,Tilav's report from IceTop \cite{this-volume}). Moreover, it agrees with extrapolations of the p- and He-spectra from direct measurements (Sec.\,\ref{sec:satellite}). However, as reported by J.\,Huang \cite{this-volume}, this finding appears in contradiction with data from Tibet-AS$\gamma$ which suggest a p-knee already at 300\,TeV so that the composition at the knee would be dominated by heavy primaries. The analysis of that data, however, is not straightforward: it combines data from different experiments and experiment phases by subtracting fluxes of different primaries, it assumes as input a heavy composition at the knee to which the data are compared \cite{Amenomori-11}, and the data selection imposes strong cuts which affect different primaries very differently. 

Only recently, KASCADE-Grande reported the observation of a steepening in the cosmic ray energy spectrum of heavy primary particles at about $8\cdot 10^{16}$\,eV \cite{KG-Fe-knee} (c.f.\,Fig.\,\ref{fig:spectra1}). This structure has a significance $3.5 \sigma$ and is seen also with smaller significance in the all-particle energy spectrum. Thus, a heavy mass composition at the second knee in the all-particle spectrum can be concluded. The position of this ``heavy knee'' in the all-particle spectrum is well in line with a rigidity scaling from the light primary knee at 3-5$\cdot 10^{15}$\,eV. Tunka and IceTop continue to improve their statistics but can already confirm a second knee just below $10^{17}$\,eV. Yet, another important finding is reported by KASCADE-Grande \cite[Haungs]{this-volume}\cite{p-ankle}: using their full data set and hard cuts on the selection of light primaries, an ankle-like feature is observed at $E \simeq 10^{17.1}$ with a significance of more than $5\sigma$ (c.f.\,Fig.\,\ref{fig:spectra1}). Confirmation of this finding would be of great importance, because the ankle-like feature in the light primaries could indicate the expected recovery of the light CR component below the ankle.

The techniques to reconstruct the mass composition of primary cosmic rays differ between experiments: KASCADE-Grande uses electron and muon numbers at ground, GRAPES in India uses muon multiplicity and $N_e$, IceTop/IceCube uses the charged particle density at ground from IceTop, $S_{\rm 125}$, and the muon density near the shower track from IceCube, $K_{\rm 70}$. Tunka uses open Cherenkov counters in which the steepness of the lateral distribution within about 120\,m from the shower core or the pulse width of the Cherenkov pulse is used to infer the composition. Despite this wide range of different observables and different atmospheric depths and despite the very different analysis techniques, the results more or less agree and provide the same overall picture. However agreeing on a quantitative level is still a long way to go.

The main obstacle in identifying Galactic CR-sources is the diffusion of CRs in the Galactic magnetic field (GMF), erasing directional information about the position of their sources. The GMF has a turbulent component that varies on scales between $l_{\rm min} \sim 1$~AU and $l_{\rm max}$ few to 200 pc. Since CRs scatter on inhomogeneities with variation scales comparable to their Larmor radius, the propagation of Galactic CRs in the GMF resembles a random walk and is well described by the diffusion approximation. Large scale anisotropies observed by the Tibet Air-Shower experiment \cite{Amenomori-06b} in the northern hemisphere for CRs at energies of a few to several hundred TeV and at angular scales of $60^\circ$ and below, thus came as a surprise. As discussed by P.\,Berghaus and D.\,Di Sciascio \cite{this-volume}, these findings, also seen by Milagro \cite{Milagro}, are now complemented by high statistics measurements from ARGO-YBJ and from IceTop in the southern hemisphere. Moreover, the structure changes with energy and appears to persist to beyond PeV energies. This anisotropy reveals a new feature of the Galactic cosmic-ray distribution, which must be incorporated into theories of the origin and propagation of cosmic rays. The origin of the anisotropies seen at different angular scales and energies is still unknown but could in principle be accounted for by specific distributions (in space and time) and individual source energy spectra of nearby recent SNRs and as such is potentially most interesting.

\subsection{The GZK Energy Region}
\label{sec:GZK}

\begin{figure}
\centering
\sidecaption
\includegraphics[width=1.05\columnwidth]{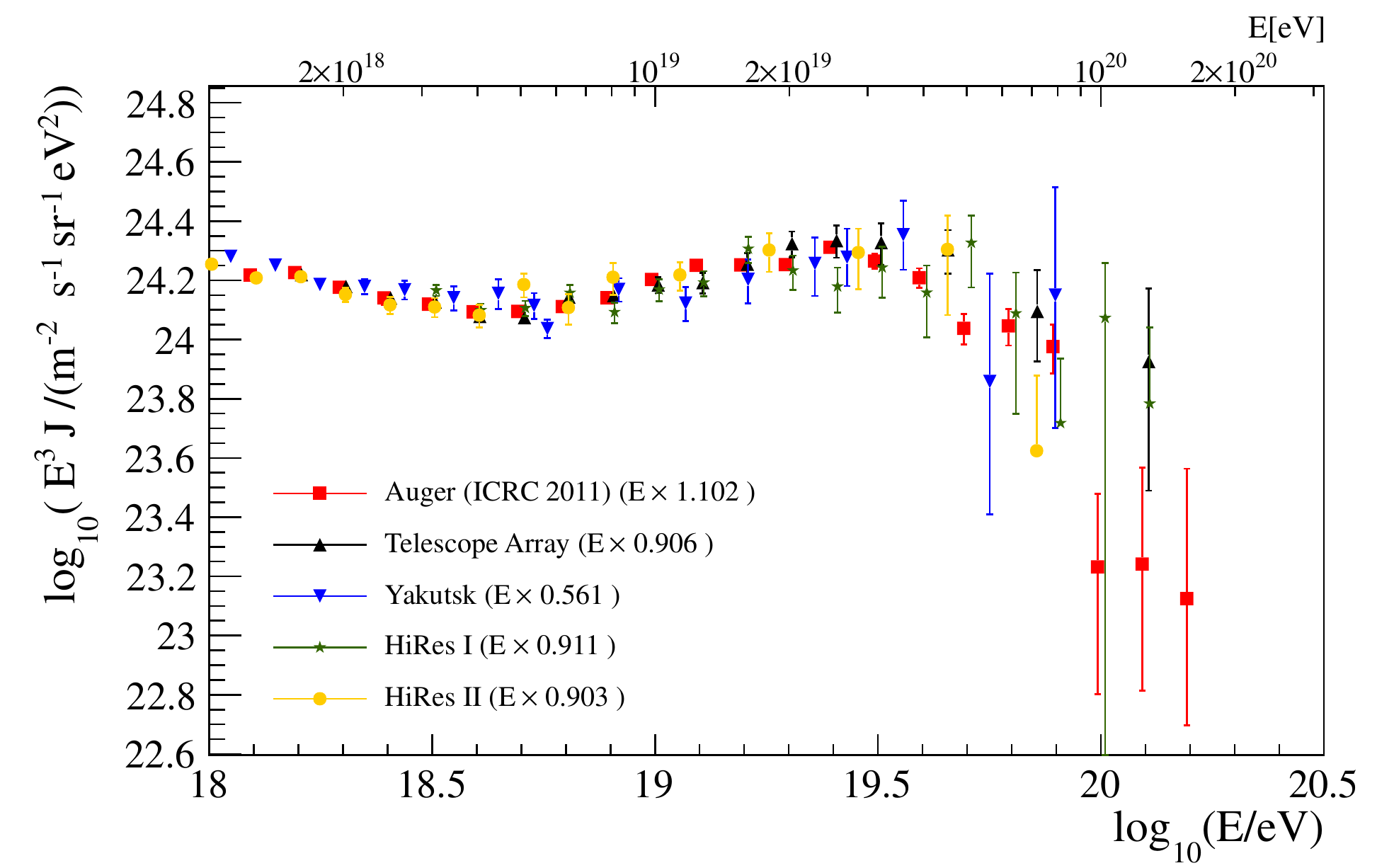}
\caption{Published energy spectra, with the flux multiplied by $E^3$, for Auger (combined Hybrid/SD), TA SD, Yakutsk SD, HiRes I, and HiRes II after energy-rescaling as shown in the figure has been applied. The reference spectrum is the average of those from Auger and TA. From \cite{E-WG} where also references to the respective data sets can be found.}
\label{fig:spectra2}
\end{figure}

At the highest energies, from the ankle to beyond $10^{20}$\,eV, the Pierre Auger Observatory \cite{auger-04,Abraham-FD-10} is the flagship in the field with an accumulated exposure of more than 30\,000\,km$^2$\,sr\,yr (see talk by L.\,Perrone \cite{this-volume}). The Telescope Array \cite{AbuZayyad:2012tr} -- presented by P.\,Sokolsky \cite{this-volume} -- due to a later start and its more than 4 times smaller area, has collected about 10 times less events. Nevertheless, interesting comparisons can already be performed and were discussed.  A detailed comparison of the energy spectra of various observatories is presented in Fig.\,\ref{fig:spectra2} \cite{E-WG}. 
As discussed at great depth in \cite{E-WG}, it is found that the energy spectra determined by the larger experiments are consistent in normalization and shape after energy scaling factors, as shown in Fig.\,\ref{fig:spectra2}, are applied. Those scaling factors are within systematic uncertainties in the energy scale quoted by the experiments. This is quite remarkable and demonstrates how well the data are understood. Nevertheless, cross-checks of photometric calibrations and atmospheric corrections have been started and as a next step, common models (e.g.\ of the fluorescence yield) should be used where possible. The data in Fig.\,\ref{fig:spectra2} clearly exhibit the ankle at $\sim 5\cdot10^{18}$\,eV and the flux suppression in accordance with the GZK-effect above $\sim 4\cdot10^{19}$\,eV.

The energy spectra by themselves, despite their high level of precision reached, do not allow one to draw conclusions on the origin of the spectral structures and thereby about the origin of CRs in the different energy regions. Additional information is obtained from the mass composition of CRs. Unfortunately, the measurement of primary masses is the most difficult task in air shower physics as such measurements rely on comparisons of data to EAS simulations with the latter serving as reference \cite{Kampert:2012hg}. Therefore, the advent of LHC data, discussed above, was of great importance to CR and EAS physics and has been awaited with great interest. 

\begin{figure}
\centering
\sidecaption
\includegraphics[width=\columnwidth]{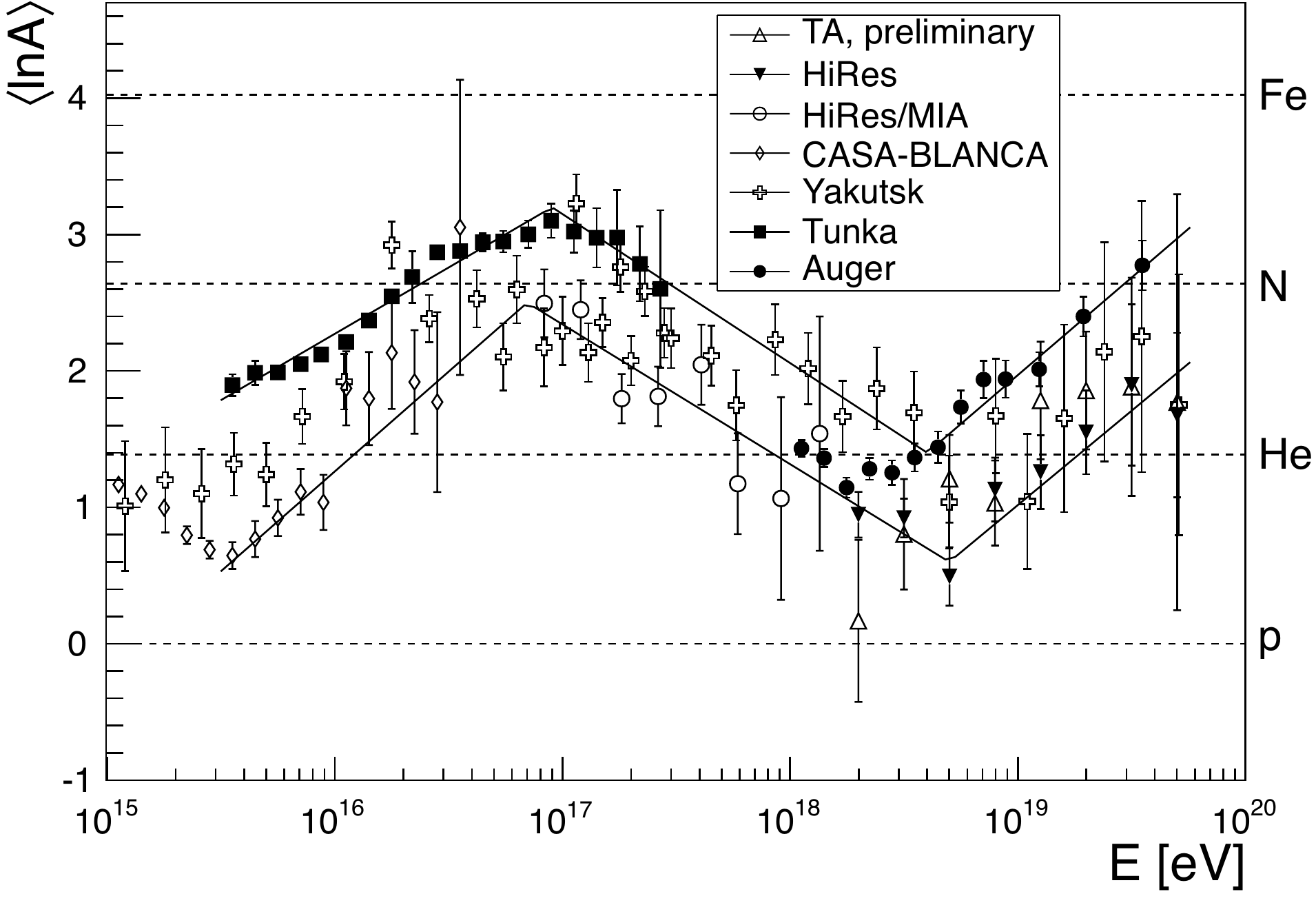}
\caption{Average logarithmic mass of cosmic rays as a function of energy derived from $X_{\rm max}$ measurements with optical detectors for the EPOS 1.99 interaction model. Lines are estimates on the experimental systematics, i.e.\ upper and lower boundaries of the data presented \cite{Kampert:2012hg}.}
\label{fig:compos}
\end{figure}

A detailed analysis of composition data from various experiments has been presented in \cite{Kampert:2012hg} with a compilation depicted in Fig.\,\ref{fig:compos}. These data complement those of the energy spectrum in a remarkable way. As can be seen, the breaks in the energy spectrum coincide with the turning points of changes in the composition: the mean mass increases above the knee, reaches a maximum at the $2^{\rm nd}$ knee, another minimum at the ankle, before it starts to rise again towards the highest energies. Different interaction models provide the same answer concerning changes in the composition but differ by their absolute values of $\langle \ln A \rangle$. It should also be noted that the rise of the mean mass at the highest energies is not without dispute (see also talk by Sokolsky \cite{this-volume}). It has been looked at in great detail in \cite{WG-compos}.
At ultra-high energies, the Auger data suggest a larger $\langle \ln A \rangle$ than all other experiments. The Auger results are consistent within systematic uncertainties with TA and Yakutsk, but
not fully consistent with HiRes. When using QGSJet-II, HiRes is compatible with the Auger data only at energies below $10^{18.5}$\,eV. When using the SIBYLL model, Auger and HiRes become compatible within a larger energy range \cite{WG-compos}.

The importance of measuring the composition at the highest energy cannot be overstated as it will be the key to answering the question about the origin of CRs at the highest energies. The same mechanism that causes the composition of Galactic CRs to increase above the knee may work also for EG-CRs above the ankle. Thereby, the break at $\sim 4\cdot10^{19}$\,eV may mark the maximum energy of EG CR-accelerators, rather than the GZK-effect. A mixture of light and intermediate/heavy primaries at the highest energies may also explain the low level of directional correlations to nearby AGN. Enhancements, presently foreseen by the Auger Collaboration will address this issue.

Two models about the putative transition from Galactic- to EG-CRs have received much attention: In the classical `ankle model' the transition is assumed to occur at the ankle. In this model, Galactic CRs above the second knee are dominated by heavy primaries before protons of EG origin start to take over and to dominate at the ankle. In the dip-model \cite{berezinsky-88}, on the other hand, the transition occurs already at the $2^{\rm nd}$ knee and is characterized by a sharp change of the composition from Galactic iron to extragalactic protons while the ankle is due to $e^+ e^-$ production of protons in the CMB. A third, `mixed composition', model has been suggested more recently \cite{Aloisio:2012tx} in which EG-CRs taking over are not considered protons but an EG mixed CR composition. Clearly, the dip-model requires a proton dominated composition essentially at all energies starting somewhat above the $2^{\rm nd}$ knee. The answer may be difficult to be given based on $\langle X_{\rm max} \rangle$ or $\langle \ln A \rangle$ alone. A much better quantity, not yet looked at, would be the RMS of these quantities, such as studied at higher energies in \cite{Abraham-xmax-10}. A rather abrupt change of composition as required by the dip-model near the $2^{\rm nd}$ knee vs a smooth change of composition as expected near the ankle in the ankle model, should become distinguishable by the RMS$(X_{\rm max})$-values already in the very near future. This has been a prime motivation for the HEAT and TALE extensions of Auger and TA, respectively.

\section{High Energy Muons and Neutrinos}
\label{sec:muons}

\begin{figure}
\centering
\sidecaption
\includegraphics[width=0.9\columnwidth]{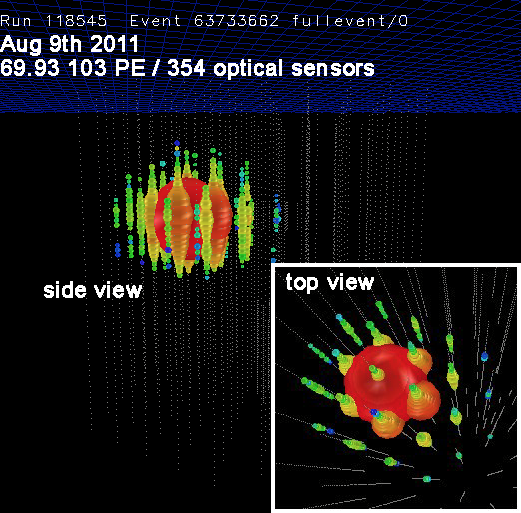}
\caption{Pictorial view of one of the two $\nu_e$ events seen in IceCube with an energy at the PeV scale \cite[Middel]{this-volume}.}
\label{fig:erni}
\end{figure}

More great news was presented from IceCube (T.\,Gaisser, E.\,Middel, P.\,Berghaus \cite{this-volume}) with the detection of two electron neutrinos with estimated energies of 1-2 PeV energy at an expected background of 0.14 events ($2.36\sigma$). The cascade like events are fully contained in the detector and one of them is shown in Fig.\,\ref{fig:erni}. 
At these energies it becomes unlikely that the events originate from conventional atmospheric neutrino fluxes, but before claiming an extraterrestrial origin, more checks are to be done.

As J.\,Brunner pointed out in his overview talk \cite{this-volume}, large high energy cosmic neutrino telescopes are now quite mature. IceCube,
for example, observes about 50,000 well-reconstructed single atmospheric neutrino events/year, with energies above 100 GeV. Although the neutrino detection probability is small, current detectors are large enough so that it is possible to study their fluxes in great detail for comparison to EAS simulations. T.\,Gaisser specifically addressed the question about signatures of enhanced prompt charm production and asked the question whether the use of
seasonal variations could help to see prompt leptons from charm decay.
Moreover, the muon charge ratio is another topic of importance for EAS simulations and due to the enormous statistics available now, there is even potential to observe the knee of the CR spectrum in the neutrino energy spectrum (see also Fedynitch and Honda). The HE neutrino telescopes may also serve the study of neutrino oscillations as was pointed out by Brunner. These are just a few examples, where high energy muons and neutrinos may provide a link to hadronic interactions and to possibly new physics.

\section{Outlook and Conclusions}
\label{sec:Summary}

Motivated by the large body of important experimental findings and new insights, the field continues to evolve very dynamically with new projects being planned or existing ones to be upgraded. In TeV $\gamma$-astronomy, reviewed by G.\,Maier, {\bf CTA} is the flagship for the next decade. In the study of low energy CRs, {\bf AMS} is the by far most complex instrument in the orbit and it has been launched on May 16, 2011. It will measure light CR isotopes from about 500~MeV to 10~GeV and is hoped to improve our understanding of CR propagation in Galaxy. First data on the positron/electron ratio have been released with more data promised to come soon.
The {\bf CALET} (CALorimetric Electron Telescope) project is another large Japanese led international mission being developed as part of the utilization plan for the International Space Station (ISS) and aims at studying details of particle propagation in the Galaxy, by a combination of energy spectrum measurements of electrons, protons and higher-charged nuclei. Similarly, {\bf Gamma-400} is planned to be flown on a Russian satellite and will combine for the first time photon and particle (electrons and nuclei) detection aiming at an angular resolution and sensitivity about 5 times better than Fermi-LAT. The idea of {\bf ISS-CREAM} is to put the CREAM detector, developed as a long duration balloon experiment, onboard the ISS, at the Japanese Experiment Modules Exposed Facility (JEM-EF) KIBO where will have good chances to reach up to $10^{15}$\,eV energy.

In Siberia, the German-Russian project {\bf HiScore} is planned to be constructed at the Tunka site. This project will use open Cherenkov counters for CR measurements around the knee and will be complemented by radio antennas to explore this new detection technology. {\bf HAWK} is being constructed in Mexico with the prime goal of studying the $\gamma$-ray sky above 100~GeV. It will also contribute to measuring CR anisotropies at TeV-energies. {\bf LHAASO}, mostly driven by the Chinese community and much larger and more complex than HAWK, serves the same scientific goals and looks forward to funding. 

At the highest energies, {\bf Auger} and {\bf TA} plan upgrades in performance and size, respectively, and started to join efforts for a Next Generation Ground-based CR Observatory {\bf NGGO}, much larger than existing experiments and aiming at good energy and mass resolution and exploring particle physics aspects at the highest energies. Four proposed and planned space missions constitute the roadmap of the space oriented community: {\bf TUS, JEM-EUSO, KLPVE, and Super-EUSO} aim at contributing step-by-step to establish this challenging field of research. They will reach very large exposures at the expense of composition measurements and particle physics capabilities. Given the resources of funding available in the next decade or two, it is unlikely that all of the above mentioned projects can be realized. Thus, priority should be given to complementarity rather than on duplication.

The cosmic ray and particle physics communities will continue to intensify their cooperations with the aim to improve the understanding of hadronic interaction models employed in EAS simulations and to exploit the potential of the most energetic cosmic rays to find signatures for new particles at an energy-scale not accessible to man-made accelerators. To establish this cooperation, the idea of an Astroparticle-Forum at CERN was discussed and supported by the community. Shortly after the Symposium, a first meeting took place at CERN to discuss issues of very forward physics and to identify the most relevant experimental observables needed to improve EAS simulations. Moreover, injecting oxygen (or nitrogen) for studying p+O (or p+N) collisions would be extremely useful for the cosmic ray community, but would also serve the study of scaling effects from pp to AA collisions at the highest energies.

\paragraph*{Acknowledgement:} 
It is a pleasure to thank the organizers of the ISVHECRI Symposium for  inviting me to this vibrant and perfectly organized conference at which many new and exciting results have been presented. I am very grateful for careful reading of the manuscript by Carola Dobrigkeit and Roger Clay. Financial support by the German Ministry of Research and Education (Grants 05A11PX1 and 05A11PXA) and by the Helmholtz Alliance for Astroparticle Physics (HAP) is gratefully acknowledged.\\[3ex]

%
%
%

\end{document}